\documentstyle[rawfonts,budapest]{article}  
\frompage{000} \topage{000}                                              

\newcommand{\be}{\begin{equation}}
\newcommand{\ee}{\end{equation}}

\newcommand{\bea}{\begin{eqnarray}}
\newcommand{\eea}{\end{eqnarray}}

\def\lag{\langle}
\def\rag{\rangle}

\def\rightmark{Spectral function of $\sigma$ meson} 
\def\leftmark{A. Patk{\'o}s et al.}
 
\title{Finite temperature spectral function of the $\sigma$ meson 
from large $N$ expansion} 
\authors{
{A. Patk{\'o}s$^{1,a}$, Zs. Sz{\'e}p$^{1,b}$ and P. Sz{\'e}pfalusy$^{2,3,c}$%
}\\[2.812mm]
{\normalsize
\hspace*{-8pt}$^1$ Department of Atomic Physics, E{\"o}tv{\"o}s University\\ 
H-1117 Budapest, Hungary\\[0.2ex] 
\hspace*{-8pt}$^2$ Research Institute for Solid State Physics and Optics\\ 
H-1525 Budapest, Hungary\\
\hspace*{-8pt}$^3$ Department of Physics of Complex Systems, 
E{\"o}tv{\"o}s University\\
H-1117 Budapest, Hungary
}}
 
\abstract{The spectral function of the scalar-isoscalar
channel  of the $O(N)$ symmetric linear $\sigma$ model is studied in the
broken symmetry phase. The investigation is  based on the leading order 
evaluation of the self-energy in the limit of large number of Goldstone 
bosons. We describe its temperature dependent variation 
in the whole low temperature phase. This variation closely reflects
the trajectory of the scalar-isoscalar quasiparticle pole. In the model with no
explicit chiral symmetry breaking we have studied near the critical point
also the corresponding dynamical exponent.
}

\keyword{spectral function, linear $\sigma$ model, large N approximation} 
\PACS{11.10Wx,12.38Mh}
 
\begin{document}
 
\maketitle
\setcounter{page}{1}

\section{Introduction}\label{intro}

The nature of the scalar meson sector of QCD, with particular emphasis on the
broad $\sigma$ meson receives increasing attention \cite{close02,tornquist01}.
This sector describes the dynamics of the order parameter of chiral symmetry, 
which is the fundamental symmetry of quantum chromodynamics, realized in
the broken symmetry phase.
Therefore the response of this multiplet to the variation of the temperature 
or the baryonic density are directly related to the 
(partial)  restoration of the chiral symmetry.

We concentrate our study on the $\sigma$-meson, therefore the $O(4)$ symmetric 
linear $\sigma$-model provides an adequate field theoretical framework. 
The finite temperature/baryon density behaviour of the $\sigma$ meson has been
studied extensively also with help of this model in the last 15 years
\cite{hatsuda01}. Since in any realistic parametrisation the
model is strongly (self)-coupled, there is no basis for the use of
the simplest perturbative techniques. 

Improved (resummed) finite temperature perturbation theory was applied
to the thermodynamics of the model \cite{chiku98} and led to the
conclusion that $\sigma$ becomes a narrow resonance with the increase 
of the temperature when its mass approaches the (temperature dependent)
two-pion threshold. The narrowing is due to the smaller phase space available
for the dominant $\sigma\rightarrow 2\pi$ decay. This situation 
manifests itself in high energy heavy ion collisions by enhanced 
two-pion production near this threshold and
the appearance of a high intensity narrow peak in the two-$\gamma$ spectra
superposed on the broad background arising from the $\pi^0$ decay.

One might remark, that results of the simplest
perturbative treatments turn out to be sensitive to the details of the
renormalisation prescription, do not reflect the so-called
hybridisation phenomenon occuring between the $\sigma$ and some
composite channels (see below), and finally, do not account correctly for  
the continuous nature of the chiral symmetry restoration.

There was some notable progress in the non-perturbative lattice determination 
of the QCD spectral functions, which did not provide to date
competitive results in the scalar-isoscalar channel 
\cite{asakawa00,wetzorke02}.

Our investigation relies on the large $N$ treatment of the thermodynamics
of the linear sigma model, where $N$ refers to the number of the Goldstone 
bosons. The leading order approximation can be shown to avoid all problems
of the usual perturbative treatment listed above. Additional credit for
the use of large $N$ techniques
 is provided by its very successful application to dynamical critical
phenomena occuring in superfluid helium \cite{kondor74,halperin77,griffin98}
and by some very promising first attempts to go beyond Hartree-type
approximations in exploring the mechanism of thermalisation for quantum fields 
\cite{berges01}. 

A more detailed report
of our results obtained in the model with no explicit $O(4)$ symmetry breaking
can be found in Ref.\cite{patkos02} and a detailed investigation of the
most realistically parametrised system with explicit symmetry breaking
is presently near completion. Here we describe the more transparent 
(though phenomenologically less appealing) case
when no explicit symmetry breaking is applied.
 
\section{Leading order large $N$ calculation of the propagators}\label{bubble}
  
The Lagrangian of the $N$-component scalar field theoretical model is 
parametrised for the large $N$ expansion in the following way:
\begin{equation}
L={1 \over
2}[\partial_\mu\phi^a\partial^\mu\phi^a-m^2\phi^a\phi^a]-{\lambda\over
24N} (\phi^a)^2(\phi^b)^2.
\end{equation}
The direction of the symmetry breaking is chosen along the $a=1$ direction
and is expressed by the shift in the corresponding field component:
\begin{equation}
\phi^a\rightarrow (\sqrt{N}\Phi+\varphi^1,\varphi^i),\qquad i=2,...,N.
\end{equation}
It is worth to note that in this parametrisation $\Phi$ is related to the
the phenomenological $f_\pi$ parameter at $N=4$ by $f_\pi =2\Phi$.

The equation of state is determined by the vanishing of the
coefficient of $\varphi^1$ in the shifted  Lagrangian:
\begin{equation}
\lag{\delta L\over \delta\varphi^1}\rag =0=\sqrt{N}\Phi \Bigl
[m^2+{\lambda \over 6}\Phi^2+{\lambda\over
6N}\lag(\varphi^a)^2\rag\Bigr ].
\label{eqstate}
\ee
To leading order in $N$ the propagator of the transversal (Goldstone) modes
receives the same tadpole contribution:
\begin{equation}
G^{-1}_{Goldstone}(p)=p^2-m^2-{\lambda\over 6}\Phi^2-{\lambda\over 6N}
\lag(\varphi^a)^2\rag,
\end{equation}
which in view of (\ref{eqstate}) leads to the masslessness of the Goldstone
modes.

The equation of state should be renormalised due to the ultraviolet divergence
of $(\lambda /N)\lag(\varphi^a)^2\rag$. To leading order in $N$ one 
retains only the contribution from the massless Goldstone modes
to the field fluctuations. 
The equation of state is rewritten with the renormalised couplings defined 
through the relations
\begin{equation}
{m^2\over\lambda}+{\Lambda^2\over 96\pi^2}=
{m^2_R\over\lambda_R},
\qquad {1\over\lambda} +{1\over 96\pi^2}\ln{e\Lambda^2\over
M_0^2}={1\over\lambda_R}, 
\label{renorm}
\end{equation}
as
\begin{equation}
m_R^2+{\lambda_R\over 6}\Phi (T)^2+{\lambda_R\over 72}T^2=0.
\end{equation} 
This expression can be rewritten as an expression for the temperature
variation of the order parameter relative to its $T=0$ value:
\begin{equation}
\Phi^2(T)=\Phi_0^2-{1\over 12}T^2,
\end{equation}
which yields for the critical temperature the prediction: $T_c^2=12\Phi_0^2=
3f_\pi^2\approx (161 {\rm MeV})^2$. This is in very good agreement with
the critical temperature determined with lattice simulations \cite{karsch01}.

The propagator of the $\sigma$ field receives to this order additional 
contribution from the infinite chain of bubble diagrams:
\begin{equation}
G^{-1}_\sigma =p^2-{\lambda\over 3}\Phi^2(T){1\over 1-\lambda b(p)/6},
\end{equation}
where $b(p)$ stands for the analytic expression of a single bubble. 
After introducing the renormalised coupling, the renormalised, $T=0$ 
expression of the bubble is given by:
\be
b_0(p)={1\over 16\pi^2}\ln{-p^2\over M_0^2},
\label{zero-bubble}
\ee
 where $M_0$ denotes the 
renormalisation scale. The $T=0$ $\sigma$-propagator looks like
\begin{equation}
G_\sigma^{-1}(p)=p^2-{\lambda_R\over 3}\Phi^2_0{1\over 1-\lambda_R\ln
(-p^2/M_0^2)/96\pi^2}.
\end{equation}

In Fig.\ref{fig1} we show the position of the $T=0$ complex
$\sigma$-pole, determined from the equation  $G_\sigma^{-1}=0$,
as a function of $\lambda_R$ in units of $f_\pi$. This
pole is a physical resonance, since it has negative imaginary part. 
It is obvious from the figure that its real part cannot exceed 
$\sim 3.5f_\pi$, while the
imaginary part gradually increases with the growth of $\lambda_R$. Choosing
$\lambda_R=310$ we find $M_\sigma/\Gamma_\sigma\sim 1$, which is close to
the phenomenologically observed value, though the mass itself is too small.

A pure imaginary pole $(p_0=iM_L)$ of the $\sigma$-propagator is found
for every 
given $T$. The theory is meaningful only for momentum scales which lie 
considerably below the tachyonic frequency. This limitates the allowed 
$\lambda_R$ range from above. In Fig.\ref{fig1} we also display the ratio 
$M_L/M_\sigma$ for several temperatures. The $T$-dependence is not very 
important therefore the limitation in the choice of $\lambda_R$ is 
nearly independent of $T$.
 
\begin{figure}[htb]
\vspace*{-0.9cm}
                 \insertplot{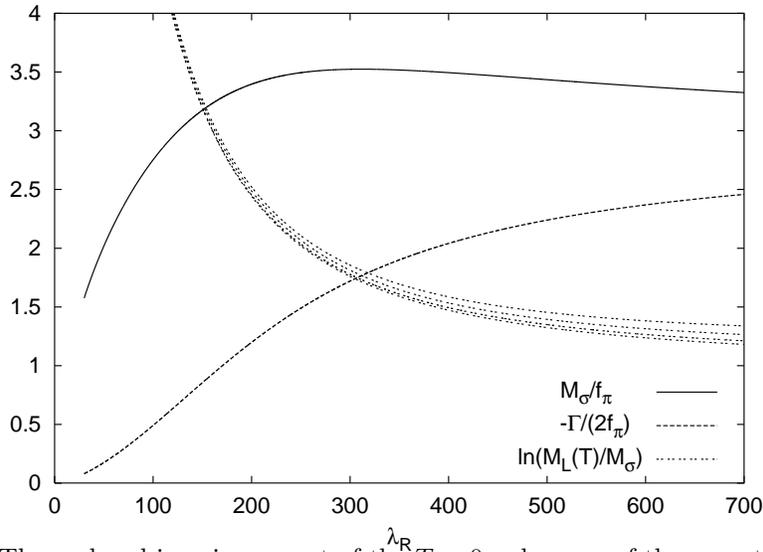}
\vspace*{-0.9cm}
\caption[]{The real and imaginary part of the $T=0$ pole mass of the 
$\sigma$ particle as a function of $\lambda_R$. We also display the location 
of the pure imaginary tachyonic pole, $M_L$.}
\label{fig1}
\end{figure}
 
\section{The scalar-isoscalar spectral function}

At finite temperature an additional term appears in the expression of
$b(p)$, which is defined in the upper halfplane of the complex $p_0$ variable:
\begin{equation}
b_T^{>}(p_0)={1\over 8\pi^2}\int_0^\infty dx{1\over e^x-1}\Bigl[{1\over
z-x}-{1\over z+x}\Bigr], \qquad z={p_0\over 2T}, \qquad {\rm Im}~p_0>0.
\end{equation}
One can compute the real and imaginary parts of the complete expression of 
$b(p)=b_0(p)+b_T(p)$ when ${\rm Im}p_0\rightarrow +0$. It determines the
spectral function in the scalar-isoscalar channel:
\be
\rho (p_0,{\bf p})=-{1\over \pi}{\rm Im}G_\sigma (p_0+i0,{\bf p}).
\ee

Similarly to $G_\sigma$ one can evaluate the propagator of the composite
field $\phi^a\phi^a-\lag\phi^a\rag^2$:
\be
F(p)={p^2b(p)/6+\Phi^2(T)/3\over
p^2(1-\lambda_Rb(p)/6)-\lambda_R\Phi^2(T)/3}. 
\ee
Obviously, its denominator is the same as that of $G_\sigma$, which reflects
the phenomenon of hybridisation.

In the two parts of Fig.\ref{figtwo} we give the spectral functions of 
$G_\sigma$ and of $F$ for ${\bf p}=0$ as functions of $p_0$ for various
temperatures. One observes that both $T=0$ curves are peaked at $p_0$ values
which agree with the real part of the $\sigma$-pole. The question we addressed
was: can one interpret the shift of the maxima for increasing temperatures 
as a reflection of the finite temperature shift of the quasi-particle pole?

In order to answer this question one has to continue the propagators 
analytically into the
lower $p_0$ halfplane (onto the second Riemann sheet) and find the poles 
smoothly joining the location of the $T=0$ pole. The expression of the
zero temperature piece $b_0(p)$ is continued in a unique manner into this
region. The harder task is the continuation of the finite temperature
contribution. One can check that if one chooses $b_T(p_0)$ to be
analytic on the positive real axis, then the relation between $b^>_T$ and
its continuation into the lower halfplane $b^<_T$ is the following:
\be
b_T^{<}(p_0)=b_T^{>}(p_0)-{i\over 4\pi}{1\over \exp
(p_0/2T)-1},\qquad {\rm Im}~p_0<0,\qquad {\rm Re}~p_0>0. 
\ee

\begin{figure}
\vspace*{-0.9cm}
\insertplot{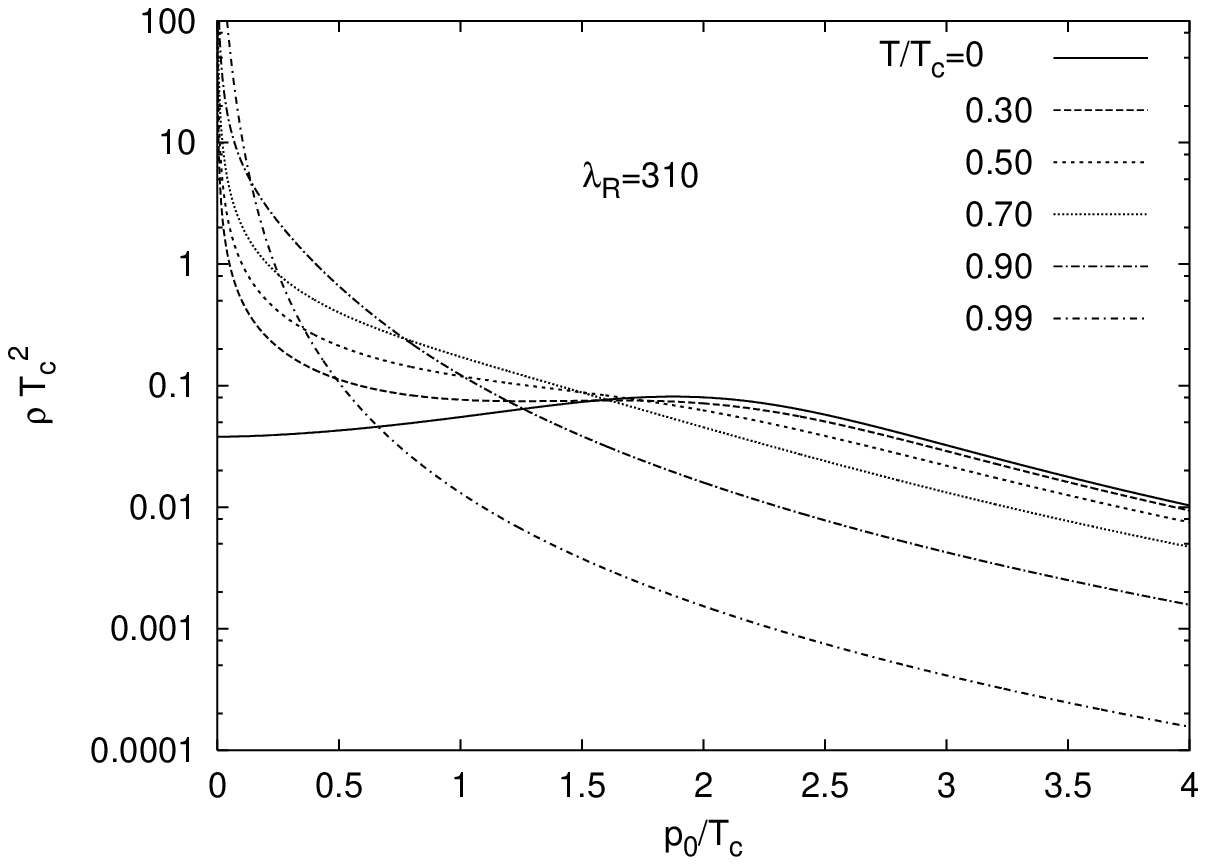}
\insertplot{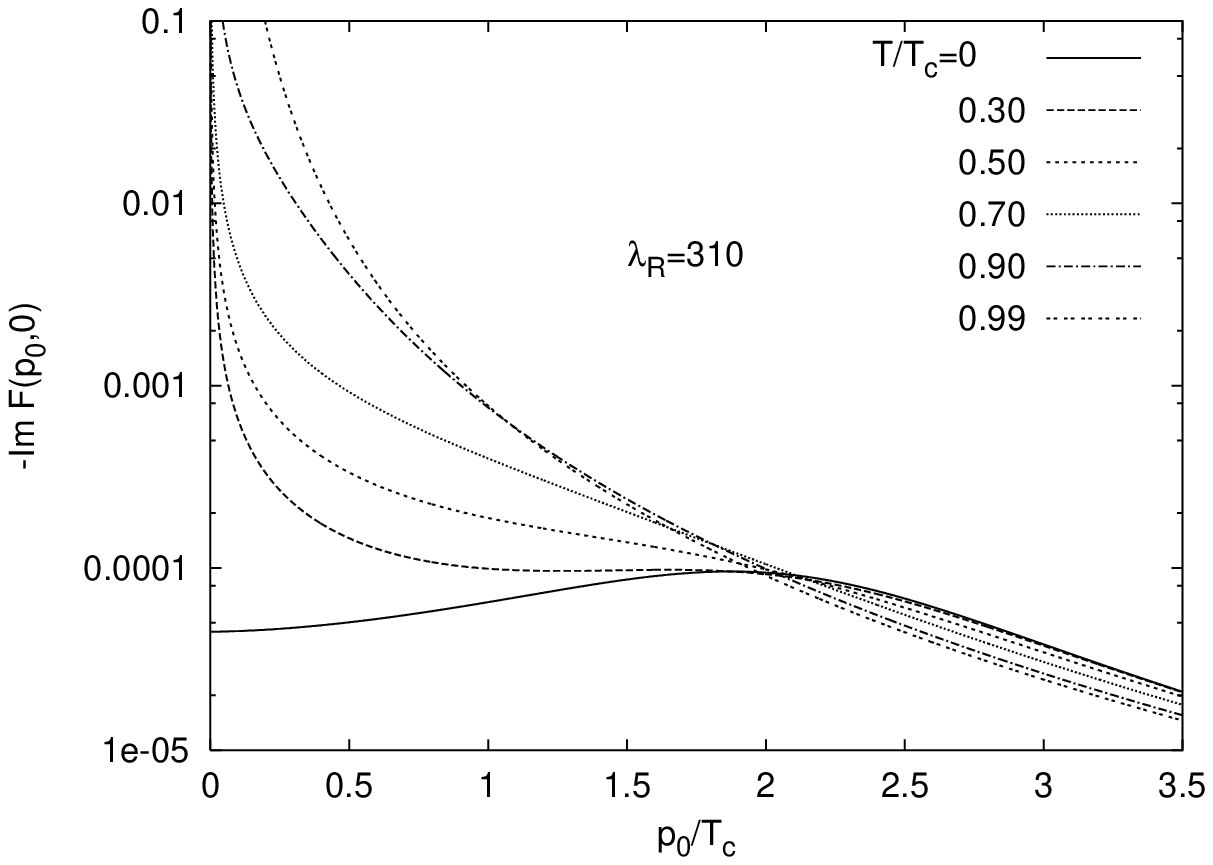}
\vspace*{-0.9cm}
\caption[]{The spectral function of the elementary and of the quadratic
composite scalar-isoscalar field as a function of the frequency $p_0$ for 
different temperatures
}
\label{figtwo}
\end{figure}

With help of this expression of $b(p_0)$ valid on the second Riemann sheet 
one can find for every $T$ the point where $G_\sigma(p_0)^{-1}$ or 
$F(p_0)^{-1}$  vanishes.

\begin{figure}[htb]
\insertplot{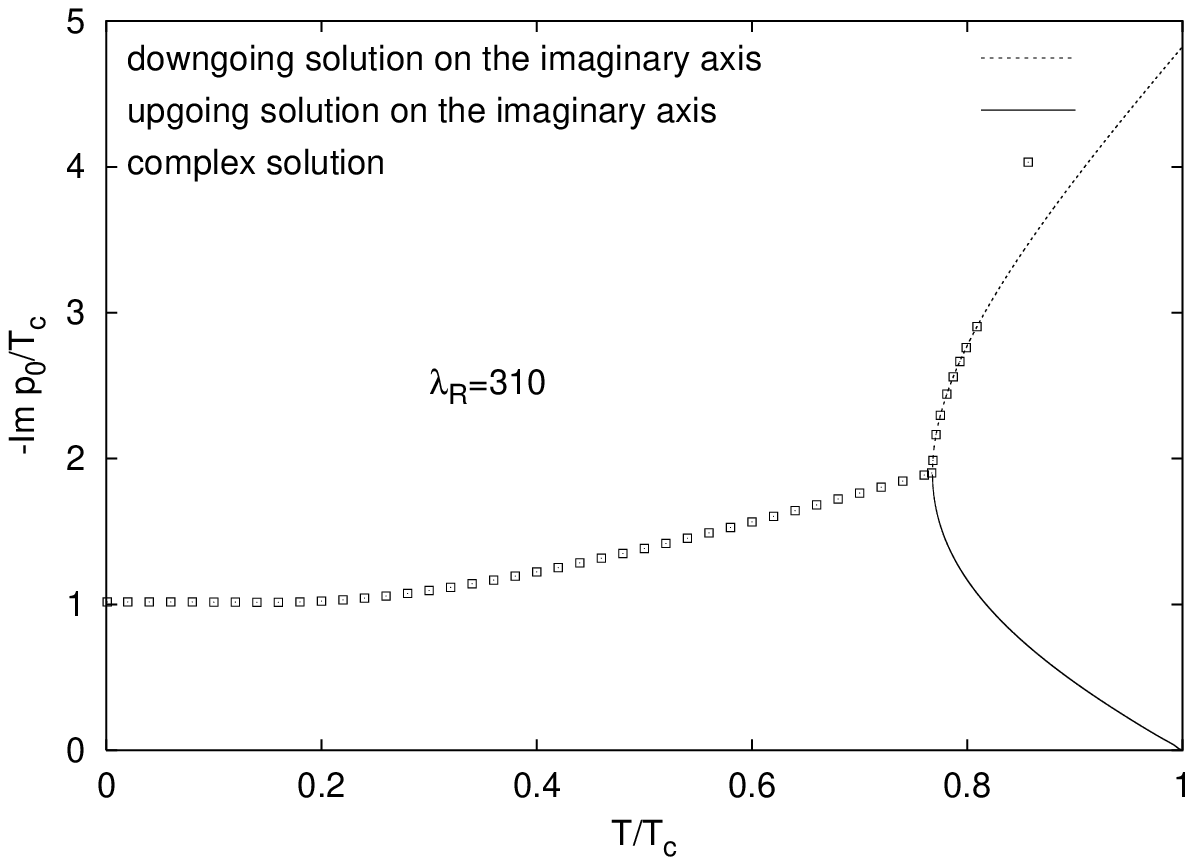}
\vspace*{-0.9cm}
\caption[]{The imaginary part of the $\sigma$-pole in the 4$^{\textrm{th}}$
quadrant of the complex $p_0$-plane including the negative imaginary axis. 
Depending on the details of the numerical algorithm used, the complex 
solution may continue into the up-going solution as well}
\label{figthree}
\end{figure}  

In Fig.\ref{figthree} we display the imaginary part of the trajectory of the 
$\sigma$-pole as a function of the temperature. Its real part gradually 
decreases with increasing temperature.
The imaginary part slightly increases at the same time. At some
$T=T_{imag}$ the pole position reaches the negative portion of the imaginary 
axis. At the same temperature a ``mirror'' pole touches the same point of 
the imaginary axis arriving from the left. The ``collision'' of the two poles
leads to the birth of two purely imaginary poles, present for higher 
temperatures. One of them moves towards, the other one away from the origin
along the imaginary axis. The narrowing of the $\sigma$ quasi-particle 
starts only for temperatures $T>T_{imag}$ and the same is true for the start of
the ``threshold enhancement'' of the spectral function.


The $O(N)$ symmetry is restored at the
temperature for which the pole moving in this direction reaches the origin. 
In the temperature
range when the poles are found on the negative imaginary axis, the 
spectral functions become cuspy. The closer moves the pole along the 
imaginary axis to the origin, the narrower the shape of the spectral function
becomes. Finally for the temperature when one of the poles arrives to the 
origin one finds analytically that
\be
\rho (p_0, T\sim T_c)\sim \delta (p_0)/p_0,
\ee
where the $\delta$-function singularity corresponds to the maximal threshold
enhancement. 

From the point of view of the spectral function the whole temperature
range $T_{imag}<T<T_c$ can be called the threshold enhancement regime, since
in this temperature interval the maximal signal comes from the direct 
neighbourhood of the imaginary axis. We believe that this is the generic
behaviour, since there is no reason to expect that from every starting
point in the $p_0$-plane the pole ``flows'' directly to the origin, without
hitting the imaginary axis.

\section{Critical region}\label{critical}

In the vicinity of the critical point, where $\xi =T/8\pi\Phi^2(T)$
is the dominant length scale and the condition $p_0,|{\bf p}|<<T$ is also
fulfilled, one can derive an equation also for the soft modes with nonzero
momentum:
\be
-3i|{\bf p}|\xi {p_0^2-{\bf p}^2\over |{\bf p}|^2}\ln{p_0-|{\bf p}|
\over p_0+|{\bf p}|}
-{1\over 4\pi}{(p_0^2-{\bf p}^2)\xi \over T_c}\ln{p_0^2-{\bf p}^2\over T_c^2}
=1.
\ee
Its solution in the approximation, when on the left hand side only the
first term is retained exhibits the form of dynamical scaling 
\cite{kondor74,halperin77}: $p_0=|{\bf p}|^{\tilde z} f(|{\bf p}|\xi)$ with 
$\tilde z=1$.
The second term provides the leading correction to scaling. In $O(N)$ models,
the dynamical exponent $z=d/2$ has been obtained for finite $N$ on the basis 
of scaling and renormalisation group arguments, where in our case $d=3$
\cite{sasvari75,sasvari74,rajagopal93}. For the correct interpretation of the
situation it is important that there are two distinct hydrodynamical regions
in the $O(N)$ model for large $N$ \cite{sasvari75}. In the true critical 
region $z=d/2$ is valid, in a precritical region $\tilde z=1-8S_d/Nd+
{\cal O}(1/N^2)$, where $S_d=2/\pi^2$ for $d=3$. The first region shrinks
when $N$ becomes large and completely disappears at $N=\infty$. 

\section{Conclusion}
The aim of this talk was to clarify the relationship of the 
temperature dependence of the spectral function characterizing the 
excitations of the chiral order parameter and the  
quasi-particle pole with the same quantum numbers (the $\sigma$-meson).
The application of the leading order analysis in a large $N$ approximation,
where $N$ is the number of Goldstone modes, was shown to avoid most of the
conceptual problems related to the application of conventional perturbative 
approaches. The pole of the $\sigma$-propagator continued analytically into
the lower $p_0$ halfplane follows a trajectory, which describes satisfactorily
the behaviour of the spectral function both near $T=0$ and in the critical 
region. The transition between the two regimes happens around $T_{imag}$, the
temperature when the pole becomes purely imaginary.

The line of analysis presented here has been applied also to the model with
explicit $O(4)$-breaking external field \cite{patkos02,patkos02a}, with
parameters chosen the closest possible to the measured characteristics of the
$\sigma - \pi$ meson system. The behaviour found in this analysis proved
generic also for the system with finite baryonic charge density. 

In view of the clear physical picture we found for the mechanism behind
the threshold enhancement of the scalar-isoscalar spectral function, it
is of interest to compute the next-to-leading corrections. This will
enable us to see to what extent can one consider the present approach also
quantitatively relevant.  

\section*{Acknowledgements}
This work has been supported by the research contract OTKA-T037689.

\section*{Notes} 
\begin{notes}
\item[a]
Speaker, E-mail: patkos@ludens.elte.hu
\item[b]
E-mail: szepzs@antonius.elte.hu
\item[c]
E-mail: psz@power.szfki.kfki.hu
\end{notes}

\vfill\eject
\end{document}